\documentclass[conference, fleqn]{IEEEtran}
\IEEEoverridecommandlockouts

\usepackage{cite}
\usepackage{graphicx}
\usepackage{multirow}
\usepackage{textcomp}
\usepackage{xcolor}
\usepackage{tabularx}
\usepackage{url}
\usepackage[ruled]{algorithm2e}

\usepackage{amsmath}  % 核心数学公式
\usepackage{newtxtext} % 完美的 Text Times 字体
\usepackage{newtxmath} % 完美的 Math Times 字体，支持标准的 \sum, + 

\definecolor{myblue}{RGB}{0, 0, 0}
\usepackage{hyperref}
\hypersetup{
    colorlinks=true,
    linkcolor=myblue,  
    citecolor=myblue,
    urlcolor=myblue 
}

\def\BibTeX{{\rm B\kern-.05em{\sc i\kern-.025em b}\kern-.08em
    T\kern-.1667em\lower.7ex\hbox{E}\kern-.125emX}}
\definecolor{authorcolor}{rgb}{0,0,0}

\begin{document}
\title{Beyond Accuracy: Physics-Informed Evaluation of Deep Learning-Based Power Flow Solvers under System Shift\\
}
\author{\IEEEauthorblockN{Dong Liu, Jochen Stiasny, Jochen Cremer, Pedro P. Vergara}
\IEEEauthorblockA{Intelligent Electrical Power Grids, Delft University of Technology, The Netherlands \\
Emails: \{D.Liu-7, J.B.Stiasny, J.L.Cremer, P.P.VergaraBarrios\}@tudelft.nl}
}

\maketitle

% Add centered IEEE copyright notice at the bottom of the first page
% \IEEEpubid{\begin{minipage}{\textwidth}\centering \vspace{1.5cm}
% 979-8-3503-9042-1/24/\$31.00~\copyright~2024 European Union
% \end{minipage}} 

\begin{abstract}
Recent advances in artificial deep learning-based power flow (PF) solvers have primarily focused on improving estimation accuracy, while their physical feasibility and operational reliability under realistic system variations remain insufficiently evaluated. This paper proposes a physics-informed evaluation framework for deep learning-based PF solvers under topology and load distribution shifts. Beyond conventional voltage-domain errors, the framework incorporates KCL gaps, branch-flow gaps, KVL residuals, continuous feasibility scores, and strict, practical, and relaxed PF feasibility rates. Four case studies are designed to investigate topology shifts, mixed topology training, load distribution shifts, and combined topology distribution shifts. Eight ANN architectures, including graph input models, load-only baselines, and physics-informed neural networks, are evaluated under consistent settings. The results demonstrate that low estimation errors do not necessarily correspond to physically feasible or operationally reliable PF solutions, particularly under combined system shifts. The proposed framework establishes an initial foundation and analysis for the development and verification of trustworthy deep learning-based PF solvers, providing support for their reliable deployment in future AI-enabled power system analysis.
\end{abstract}


\section{Introduction}

Deep learning-based power flow (PF) solvers have attracted increasing attention as fast data-driven alternatives to conventional model-based PF methods. These solvers are particularly attractive because they require less detailed topology information and are relatively robust to noisy, erroneous, or missing measurements~\cite{stiasny2026residual}. However, most existing deep learning-based PF solvers are primarily developed and evaluated based on accuracy, whereas their practical reliability in real grid analysis remains insufficiently examined. In particular, accurate estimates of voltage magnitudes and phase angles, or even values within acceptable limits, do not necessarily guarantee a physically feasible or secure operating state, since the corresponding solution may still violate operational constraints such as line PF limits. Therefore, the lack of systematic verification of physical consistency and operational feasibility under realistic system changes may introduce additional reliability risks and hinder the practical deployment of deep learning-based PF solvers~\cite{11134748}.

Basic deep learning-based PF solvers primarily focus on learning an accurate input–output mapping, where ANNs are trained on historical datasets and then used to rapidly estimate DC PF results~\cite{9772026}. Physics-informed neural network (PINN)-based PF solvers have further been developed by embedding simplified admittance matrices, thereby improving calculation accuracy and robustness~\cite{kaseb2026physics}. For large scale power grids, PowerFlowNet~\cite{lin2024powerflownet} employs graph neural networks (GNNs) by representing power systems as graphs and achieves performance comparable to the traditional Newton--Raphson method. Similarly, a model-driven graph convolutional neural network is developed based on a simplified linear power flow model to enhance both interpretability and accuracy~\cite{10176302}. The application of deep learning-based PF solvers in distribution networks is investigated in~\cite{10444519}, where three deep neural networks are proposed to approximate PF results in three phase unbalanced distribution networks. For unbalanced distribution networks, the PowerFlowMultiNet is introduced as a multigraph GNN with separate phase-wise graphs to capture phase-specific operating characteristics, while a message-passing strategy is adopted to model time-dependent features~\cite{10684547}. To further enhance deep learning-based AC PF solvers for distribution networks, line losses are incorporated into the loss function in~\cite{10863674}, and Kirchhoff’s laws and active power balance constraints are incorporated into the training process, which also reduces training time~\cite{kim2026physics}. In addition, considering the growing importance of data privacy, a privacy-preserving deep learning-based PF method is proposed to maintain calculation accuracy while protecting sensitive data~\cite{11098734}. Nevertheless, existing deep learning-based PF solvers mainly emphasise prediction accuracy and the incorporation of physical constraints during training or in the loss function. %  To improve the accuracy of ANN-based PF solvers in low-voltage distribution networks,

However, low estimation errors during training do not guarantee a physically feasible or secure operating point. Since line power flows, bus injections, and network losses are derived from voltage magnitudes and phase angles together with topology and line parameters~\cite{11134748}, small voltage errors may propagate to other datasets. Thus, even if the estimated voltage magnitudes and phase angles remain within operational limits, the corresponding operating state may still violate Kirchhoff’s laws, produce inaccurate line power flows, or mask line overloads~\cite{9894104}. This limitation may lead to misleading conclusions in PF-based analysis, such as power balance verification, branch-loading assessment, or security screening~\cite{kaseb2026physics}. Moreover, many deep learning-based PF solvers are trained and evaluated under fixed network topology and fixed data distributions. In practical power grids, however, topology and operating conditions frequently change due to switching actions, distributed generation variability, and demand fluctuations~\cite{10684754}. Consequently, a systematic physics-informed evaluation is required to assess whether deep learning-based PF solvers can preserve physical consistency and operational feasibility when the estimated variables are used to compute nodal injections, branch flows, and security constraints under topology and load distribution shifts.

To address this gap, this paper proposes a physics-informed evaluation framework for deep learning-based PF solvers under topology and load distribution shifts. A unified IEEE 14-bus dataset with 50 topology scenarios and multiple load distribution scenarios is used to evaluate eight ANN architectures, including graph input models, load-only baselines, and physics-informed neural networks. Beyond conventional prediction errors, the framework incorporates relative KCL gaps, branch flow gaps, KVL residuals, continuous feasibility scores, and strict, practical, and relaxed PF feasibility rates. Four case studies investigate topology shifts, mixed-topology training, load distribution shifts, and combined topology distribution shifts. The framework enables systematic assessment of physical consistency and operational feasibility beyond statistical accuracy, providing a basis for the verification and reliable deployment of trustworthy deep learning-based PF solvers under realistic system variations.

\section{Physics-Informed Evaluation Framework}
\begin{figure}
\centering
\includegraphics[width=0.5\textwidth,height=0.28\textwidth]{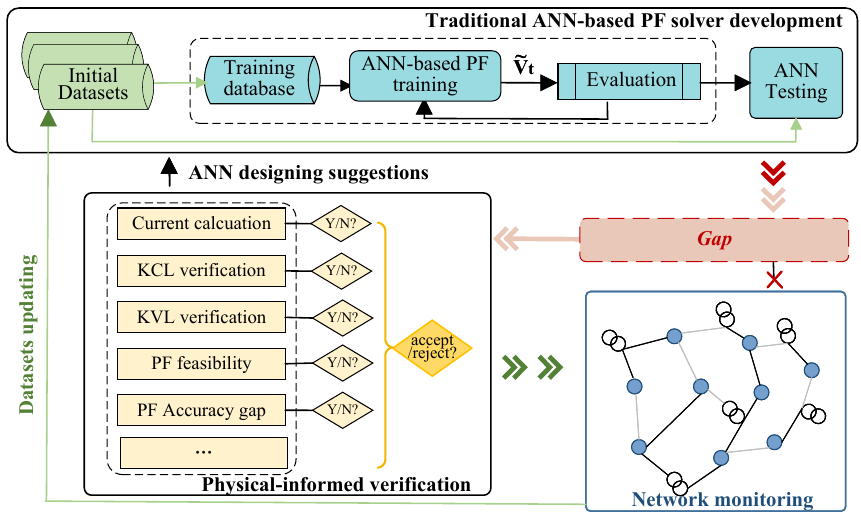}
\vspace{-0.5cm}
\caption{Physics-informed evaluation framework for deep learning-based PF solver.}
\vspace{-0.5cm}
\label{fig1}
\end{figure}

The proposed framework is illustrated in \textcolor{myblue}{Fig.~\ref{fig1}}. It consists of three main stages: (i) generation of topology- and distribution-aware PF datasets, (ii) training of multiple deep learning-based PF solvers under different input-output settings, and (iii) physics-informed evaluation of the estimated outputs. All ANN models are trained to estimate the outputs, whereas the feasibility of the predicted solution is evaluated afterwards using AC PF equations, KCL, KVL, and branch-flow consistency.

\subsection{Deep Learning-Based PF Calculation}

The deep learning-based PF solver is considered as a generic surrogate model without assuming a specific ANN architecture. Let $\mathbf{x}$ denote the available input, which may be represented as a load vector, a node-feature matrix, or a graph with branch features depending on the model class. The learning target is defined as the set of unknown PF variables $\hat{\mathbf{z}}$. Specifically, for PQ buses, both voltage components are unknown, whereas for PV buses, the voltage magnitude is specified and only the remaining unknown variables need to be estimated. Let $f_{\theta}\left(\cdot\right)$ denote the ANN surrogate parameterized by $\theta$.
\begin{equation}
    \hat{\mathbf{z}}
    =    f_{\theta}\left(\mathbf{x}\right)
\end{equation}

Based on $\hat{\mathbf{z}}$, the complete complex voltage vector $\hat{\mathbf{V}}$ is reconstructed by combining the predicted states with the known voltage references at slack and PV buses. In distribution network applications, where most non-slack buses are commonly modeled as PQ buses (i.e., household load datasets), $\mathbf{x}$ typically contains the load conditions of all PQ buses, while the ANN output corresponds to the voltage magnitude of all non-slack buses.

% In rectangular coordinates, this reconstruction can be generally written as
% \begin{equation}
%     \hat{V}_i =
%     \begin{cases}
%     \hat{\mu}_i+j\hat{\omega}_i, & i\in\mathcal{N}_{\mathrm{PQ}},\\
%     V_i^{\mathrm{ref}}(\hat{\delta}_i), & i\in\mathcal{N}_{\mathrm{PV}},\\
%     V_i^{\mathrm{ref}}, & i\in\mathcal{N}_{\mathrm{slack}},
%     \end{cases}
%     \label{eq:ann_voltage_reconstruction}
% \end{equation}
% where $\mathcal{N}_{\mathrm{PQ}}$, $\mathcal{N}_{\mathrm{PV}}$, and $\mathcal{N}_{\mathrm{slack}}$ denote the sets of PQ, PV, and slack buses, respectively. For PQ buses, $\hat{\mu}_i$ and $\hat{\omega}_i$ are the predicted real and imaginary voltage components. For PV buses, $V_i^{\mathrm{ref}}(\hat{\delta}_i)$ denotes the complex voltage reconstructed using the specified voltage magnitude and the estimated voltage angle $\hat{\delta}_i$. The slack-bus voltage is directly assigned from its reference value.

After reconstructing $\hat{\mathbf{V}}$, the variables used for evaluation, including line power flows, KCL residuals, and KVL residuals, are derived from $\hat{\mathbf{V}}$ and the corresponding network parameters according to the AC network equations.

\subsection{Prediction Error-Based Evaluation Metric}

A conventional way to evaluate deep learning-based PF solvers is to measure the error between the estimated and true variables. The mean square error (MSE) based loss function is normally defined as:
\begin{equation}
    \mathcal{L}_{MSE} =
    \frac{1}{N_sN_b}  \sum_{s\in\mathcal{N}_s}\sum_{i\in\mathcal{N}_b}
    \left[
    (\hat{\mu}_{i}^{s}-\mu_{i}^{s})^2+
    (\hat{\omega}_{i}^{s}-\omega_{i}^{s})^2
    \right]
    \label{eq:state_mse}
\end{equation}
where $N_s$ is the number of test samples, and $\mathcal{N}_b$ denotes the set of buses whose voltage-related state variables are estimated by the ANN. For PQ buses, both real and imaginary voltage components  $\mu_{i}^{s}$ and $\omega_{i}^{s}$ are estimated, while for PV buses, only the unknown voltage-related variables are included according to the adopted state representation.

For PINNs, the training objective may further include a physical-consistency term in addition to the loss function.
\begin{equation}
    \mathcal{L}_{\mathrm{PINN}}
    =
    \mathcal{L}_{\mathrm{sup}}   + W_{\mathrm{phy}}\mathcal{L}_{\mathrm{phy}}
    \label{eq:pinn_loss}
\end{equation}
where $\mathcal{L}_{\mathrm{sup}}$ denotes the supervised estimation loss (e.g., $\mathcal{L}_{MSE}$), $\mathcal{L}_{\mathrm{phy}}$ represents the penalty associated with physical constraints, and $W_{\mathrm{phy}}$ is the weight to guide the convergence of training. Although these metrics quantify estimation accuracy and may encourage physical consistency during training, they do not directly determine whether the reconstructed operating state is feasible or secure.

\subsection{Physics-Informed Evaluation Metrics}

Given an obtained voltage vector $\hat{\mathbf{V}}$, the corresponding bus complex injection is computed from the bus admittance matrix $\mathbf{Y}$ as:
\begin{equation}
    \hat{S}_i =
    \hat{V}_i
    \left(\sum_{j=1}^{N}Y_{ij}\hat{V}_j\right)^*,
    \label{eq:nodal_injection}
\end{equation}
where $(\cdot)^*$ denotes complex conjugation and $N$ is the number of buses. $ij$ are the indices of the lines in $\mathcal{L}_l$. Based on $\hat{\mathbf{V}}$, several physics-informed metrics are introduced to evaluate whether the estimated voltage states are not only numerically accurate but also physically consistent and operationally feasible.

\subsubsection{KCL Gap}

The first metric evaluates the consistency of the predicted state with KCL at PQ buses. The KCL residual index is defined as:
\begin{equation}
    \Phi_{\mathrm{KCL}} =
    \frac{1}{N_s|\mathcal{P}|}
    \sum_{s=1}^{N_s}\sum_{i\in\mathcal{P}}
    |\hat{S}_{i}^{(s)}-S_i^{\star(s)}|^2 ,
    \label{eq:kcl_index}
\end{equation}
where $\mathcal{P}$ is the set of PQ buses, and $S_i^\star$ denotes the reference complex injection obtained from the PF solution. A smaller $\Phi_{\mathrm{KCL}}$ indicates better nodal power balance consistency.

\begin{figure*}[!t]
\centering
\includegraphics[width=0.98\textwidth,height=0.33\textwidth]{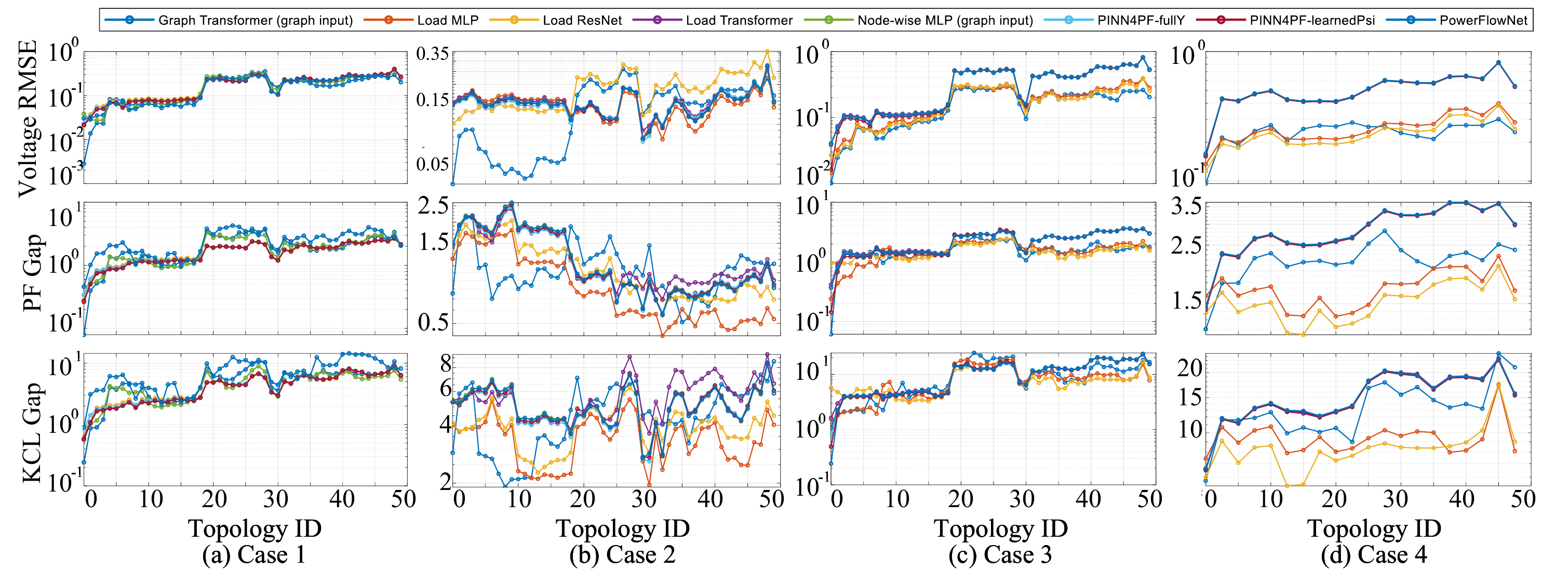}
\vspace{-0.2cm}
\caption{Comparison of voltage RMSE, relative KCL gap $\Delta_{\mathrm{KCL}}$, and relative PF gap $\Delta_{\mathrm{PF}}$ for deep learning-based PF solvers across the four case studies.}
\label{fig2}
\vspace{-0.2cm}
\end{figure*}

Since the absolute residual depends on the system loading level, a relative KCL gap is further reported as:
\begin{equation}
    \Delta_{\mathrm{KCL}} =
    \sqrt{
    \frac{
    \frac{1}{|\mathcal{P}|}\sum_{i\in\mathcal{P}}
    |\hat{S}_i-S_i^\star|^2
    }{
    \frac{1}{|\mathcal{P}|}\sum_{i\in\mathcal{P}}
    |S_i^{\mathrm{load}}|^2+\epsilon
    }},
    \label{eq:kcl_gap}
\end{equation}
where $S_i^{\mathrm{load}}$ is the specified complex load and $\epsilon$ is a small constant introduced for numerical stability.

\subsubsection{PF Gap}

To assess line power flow consistency, the predicted complex power flow is computed for both directions of each line $(i,j)\in\mathcal{E}$ as:
\begin{equation}
    \hat{S}_{ij}=
    \hat{V}_i\left(y_{ij}(\hat{V}_i-\hat{V}_j)\right)^* ,
    \label{eq:branch_flow}
\end{equation}
and analogously for $\hat{S}_{ji}$. Here, $y_{ij}$ is the admittance of line $(i,j)$. The line PF deviation index is then defined as
\begin{equation}
    \Phi_{\mathrm{BF}} =
    \frac{1}{2N_s|\mathcal{\mathcal{L}_l}|}
    \sum_{s=1}^{N_s}\sum_{(i,j)\in\mathcal{\mathcal{L}_l}}
    \left(
    |\hat{S}_{ij}^{(s)}-S_{ij}^{\star(s)}|^2+
    |\hat{S}_{ji}^{(s)}-S_{ji}^{\star(s)}|^2
    \right),
    \label{eq:branch_index}
\end{equation}
where $S_{ij}^{\star}$ and $S_{ji}^{\star}$ denote the reference line PF values in the two directions. This metric quantifies the deviation between the line PF implied by the predicted voltage state and that obtained from the reference PF solution.

To make the line PF deviation comparable across loading conditions, the relative PF gap is defined as:
\begin{equation}
    \Delta_{\mathrm{PF}} =
    \sqrt{
    \frac{\Phi_{\mathrm{BF}}}
    {\max\{\sigma_{\mathrm{flow}},\sigma_{\mathrm{load}}\}+\epsilon}},
    \label{eq:pf_gap}
\end{equation}
% where $\sigma_{\mathrm{flow}}$ and $\sigma_{\mathrm{load}}$ denote the average line PF scale and load scale, respectively.

where $\sigma_{\mathrm{flow}}$ is obtained by averaging the squared
magnitudes of the reference PF in both directions across all
lines and samples, while $\sigma_{\mathrm{load}}$ is obtained by
averaging the squared magnitudes of the specified complex load powers
across all PQ buses and samples.

\subsubsection{KVL Cycle Residual Index}

KVL is evaluated over a set of independent cycles $\mathcal{C}$. For each cycle $c\in\mathcal{C}$, the cycle residual is computed as:
\begin{equation}
    r_{\mathrm{KVL}}(c)=
    \left|
    \sum_{(i,j)\in c}
    (\hat{V}_i-\hat{V}_j)
    \right|.
    \label{eq:kvl_residual}
\end{equation}

The KVL cycle residual index is defined as:
\begin{equation}
    \Phi_{\mathrm{KVL}} =
    \frac{1}{N_s|\mathcal{C}|}
    \sum_{s=1}^{N_s}\sum_{c\in\mathcal{C}}
    r_{\mathrm{KVL}}^{(s)}(c).
    \label{eq:kvl_index}
\end{equation}

This metric measures whether the reconstructed voltage state satisfies voltage consistency around network cycles.

\subsubsection{PF Feasibility Rate}
\begin{table*}[t]
\caption{CASE DESCRIPTIONS AND THEIR OBJECTIVE QUESTIONS.}    \centering
    \small
    \renewcommand{\arraystretch}{1.1}
    \begin{tabularx}{\textwidth}{p{2.4cm}|X|X|X|X}
        \hline
        \textbf{Case name} 
        & \textbf{Case 1: Topology shift only} 
        & \textbf{Case 2: Fully mixed training upper bound} 
        & \textbf{Case 3: Distribution shift only / mixed} 
        & \textbf{Case 4: Topology + distribution double shift} \\
        \hline
        
        \textbf{Data split} 
        & \begin{tabular}[t]{@{}l@{}}
          Train: raw0\\
          Test: raw0--raw49
          \end{tabular}
        & \begin{tabular}[t]{@{}l@{}}
          Train/Val/Test:\\
          split from merged datasets
          \end{tabular}
        & \begin{tabular}[t]{@{}l@{}}
          Train: raw0\_id\\
          Test: raw0--raw49 $\times$\\
          data distribution shifts
          \end{tabular}
        & \begin{tabular}[t]{@{}l@{}}
          Train: raw0\_id\\
          Test: raw30--raw49 $\times$\\
         data distribution shifts
          \end{tabular} \\
        \hline
        
        \textbf{Objective question} 
        & How robust are the solvers to unseen topology shifts under similar load distributions?
        & What is the performance when topology diversity is included during training?
        & How robust are the solvers to load distribution shifts under a similar topology?
        & How reliably do the solvers perform under simultaneous topology and load distribution shifts? \\
        \hline
    \end{tabularx}
    \label{table1}
    \vspace{-0.5cm}
\end{table*}
In addition to residual-based metrics, feasibility indicators are introduced to determine whether each estimated operating point satisfies a prescribed set of physical and operational criteria.
% \begin{align}
%     \Delta_{\mathrm{KCL}} \leq \tau_{\mathrm{KCL}}^\ell, \quad & \Phi_{\mathrm{KVL}} \leq \tau_{\mathrm{KVL}},\nonumber\\
%     \Delta_{\mathrm{PF}} \leq \tau_{\mathrm{PF}}^\ell, \quad & \underline{V} \leq |\hat{V}_i| \leq \overline{V}, \quad \forall i \in \mathcal{N}. \label{eq:feasibility_condition_2}
% \end{align}
The feasibility rate at level $\ell$ is defined as:
\begin{equation}
    \begin{split}
        R_{\mathrm{feas}}^\ell = \frac{1}{N_s} \sum_{s=1}^{N_s} \mathbb{I} \Biggl\{ 
        &\Delta_{\mathrm{KCL}}^{(s)} \leq \tau_{\mathrm{KCL}}^\ell, \,
        \Delta_{\mathrm{PF}}^{(s)} \leq \tau_{\mathrm{PF}}^\ell, \,\\
        & \Phi_{\mathrm{KVL}}^{(s)} \leq \tau_{\mathrm{KVL}}, \\
        &\underline{V} \leq |\hat{V}_i^{(s)}| \leq \overline{V}, \; \forall i \in \mathcal{N} \Biggr\},
    \end{split}
    \label{eq:feas_rate}
\end{equation}
where $\mathbb{I}\{\cdot\}$ is the indicator function and
$\ell \in \{\mathit{strict},\,\mathit{prac},\,\mathit{relax}\}$ denotes
the feasibility level. Since conventional PF references are obtained using iterative solvers with finite convergence tolerances, deep learning-based PF solutions are assessed using residual thresholds rather than exact equality. The strict level represents near-solver-level accuracy, the practical level provides the main engineering feasibility criterion, and the relaxed level identifies near-feasible predictions. These thresholds are selected according to the numerical accuracy of the reference solver and the operational requirements of the power system.

% In the implementation, the voltage interval is set to $[0.85,1.15]$ p.u. and $\tau_{\mathrm{KVL}}=10^{-5}$. The KCL/PF thresholds are set to $(0.10,0.10)$ for the strict level, $(2.00,1.00)$ for the practical level, and $(8.00,3.00)$ for the relaxed level.

\begin{table}[!t]
\caption{DEEP LEARNING-BASED PF SOLVERS SETTING.}
\centering
    \scriptsize
    \renewcommand{\arraystretch}{1.10}
    \begin{tabular}{p{2.15cm}|p{5.65cm}}
        \hline
        \textbf{Model} & \textbf{Main setting} \\
        \hline
        PowerFlowNet \cite{lin2024powerflownet} 
        & Graph input; 2 GraphConv layers; hidden 64; MSE loss. \\
        \hline
        
        Graph Transformer (graph input)
        & Graph input; 2 transformer encoder layers; 4 heads; hidden 64; MSE loss. \\
        \hline
        
        Node-wise MLP (graph input) 
        & Graph node features; shared MLP 64--64; no message passing; MSE loss. \\
        \hline
        
        PINN4PF-fullY 
        & Load input; MLP 64--64 + two 32-unit heads; supervised + full-$\mathbf{Y}$ physics loss. \\
        \hline
        
        PINN4PF-learnedPsi \cite{kaseb2026physics}
        & Load input; same voltage heads + learned $\Psi$ heads; supervised + learned-$\Psi$ physics loss. \\
        \hline
        
        Load MLP  
        & Load input; feed-forward MLP 64--64; supervised voltage loss. \\
        \hline
        
        Load Transformer 
        & Load tokens; 2 transformer encoder layers; 4 heads; hidden 64; supervised loss. \\
        \hline
        
        Load ResNet 
        & Load input; residual MLP; hidden 128; supervised voltage loss. \\
        \hline
    \end{tabular}

    \label{table2}
    \vspace{-0.5cm}
\end{table}

\subsubsection{Continuous Feasibility Score}

Binary feasibility rates may become uninformative when no samples satisfy the strict or practical criteria. Therefore, a continuous feasibility score is further introduced:
\begin{equation}
    \Gamma =
    \frac{1}{1+\Delta_{\mathrm{KCL}}/\tau_{\mathrm{KCL}}^{\mathrm{prac}}}
    \frac{1}{1+\Delta_{\mathrm{PF}}/\tau_{\mathrm{PF}}^{\mathrm{prac}}}
    \exp(-\rho_V),
    \label{eq:feas_score}
\end{equation}
where $\rho_V$ penalizes violations of the voltage interval. A larger $\Gamma$ indicates a more physically consistent and operationally feasible prediction. Unlike binary feasibility rates, $\Gamma$ provides a graded measure of feasibility and remains informative even when $R_{\mathrm{feas}}^{\mathrm{strict}}$ or $R_{\mathrm{feas}}^{\mathrm{prac}}$ is zero.

In the implementation, the voltage penalty is computed from the maximum lower- and upper-bound violations:
\begin{equation}
    \rho_V =
    \frac{
    \max_i[\underline{V}-|\hat{V}_i|]_+
    +
    \max_i[|\hat{V}_i|-\overline{V}]_+
    }{0.05},
\end{equation}
where $[x]_+=\max(x,0)$.

The exported metric set therefore includes voltage MSE, KCL residual index $\Phi_{\mathrm{KCL}}$, relative KCL gap $\Delta_{\mathrm{KCL}}$, relative PF gap $\Delta_{\mathrm{PF}}$, KVL cycle residual index $\Phi_{\mathrm{KVL}}$, strict/practical/relaxed feasibility rates $R_{\mathrm{feas}}^\ell$, continuous feasibility score $\Gamma$, and feasible/infeasible sample counts.

\section{Case Study}
\subsection{Dataset and Experimental Settings}

The dataset is generated from the IEEE 14-bus system. The bus set is fixed, while the branch set is modified to construct 50 topology scenarios, denoted by $\mathcal{T}=\{0,\ldots,49\}$. The original topology is denoted as raw0, and the remaining scenarios are obtained by progressively changing branch connectivity while maintaining network connectivity and AC PF solvability. Thus, the topology index is used as a controlled indicator of topology variation.

For each topology, AC PF simulations are performed to obtain bus-voltage states, bus injections, and branch power flows. Each sample contains load-vector inputs for load-only models and topology-related graph features for topology-aware models. To evaluate load distribution shifts, each topology is further combined with several load scenarios, including \emph{id}, \emph{high}, \emph{low}, \emph{wide}, \emph{skewed}, and \emph{correlated}. These scenarios modify the load level, variance, spatial imbalance, or cross-bus correlation. The dataset for topology $t$ and distribution $d$ is denoted by raw$t$\_$d$.

\begin{table}[!t]
\caption{THRESHOLD SETTINGS USED IN THE CASE STUDIES.}    \centering
    \small
    \renewcommand{\arraystretch}{1.15}
    \begin{tabular}{p{2.8cm}|p{4.7cm}}
        \hline
        \textbf{Indicator} & \textbf{Setting} \\
        \hline
        Voltage interval 
        & $0.85 \leq |\hat{V}_i| \leq 1.15$ p.u. \\
        \hline
        KVL tolerance 
        & $\Phi_{\mathrm{KVL}} \leq 10^{-5}$ \\
        \hline
        Strict feasibility 
        & $\Delta_{\mathrm{KCL}}\leq0.10$, $\Delta_{\mathrm{PF}}\leq0.10$ \\
        \hline
        Practical feasibility 
        & $\Delta_{\mathrm{KCL}}\leq2.00$, $\Delta_{\mathrm{PF}}\leq1.00$ \\
        \hline
        Relaxed feasibility 
        & $\Delta_{\mathrm{KCL}}\leq8.00$, $\Delta_{\mathrm{PF}}\leq3.00$ \\
        \hline
        Feasibility condition
        & KCL, PF, KVL, and voltage-interval constraints must all be satisfied. \\
        \hline
        Feasibility score 
        & Continuous score based on practical KCL/PF thresholds and voltage limit violations. \\
        \hline
    \end{tabular}
    \label{table3}
    \vspace{-0.5cm}
\end{table}

Four case studies are designed to evaluate topology shift, mixed-topology training, load distribution shift, and simultaneous topology--distribution shift, as summarized in Table~\ref{table1}. The evaluated ANN architectures and feasibility thresholds are listed in Tables~\ref{table2} and~\ref{table3}, respectively. For each case, the training dataset is split into 50\% training, 20\% validation, and 30\% test samples, while target scenarios are evaluated using all available samples. In the mixed-topology setting, samples from raw0--raw49 are shuffled and split, with topology labels retained for post-evaluation analysis.

All models are trained for 300 epochs with a mini-batch size of 128. Graph-input models use AdamW with a learning rate of $10^{-3}$, while load-only and PINN-based models use Adam with a learning rate of $10^{-3}$ and StepLR$(100,0.5)$. The PINN loss is formulated as:
\begin{equation}
    \mathcal{L}_{\mathrm{PINN}}
    =
    \beta_0 \mathcal{L}_{\mathrm{sup}}
    +
    \beta_1 \mathcal{L}_{\mathrm{phys}},
\end{equation}
where $\beta_0=1-\beta_1$. The physics weight $\beta_1$ is gradually increased during training and capped at 0.71.

\begin{figure}
\centering
\includegraphics[width=0.5\textwidth,height=0.5\textwidth]{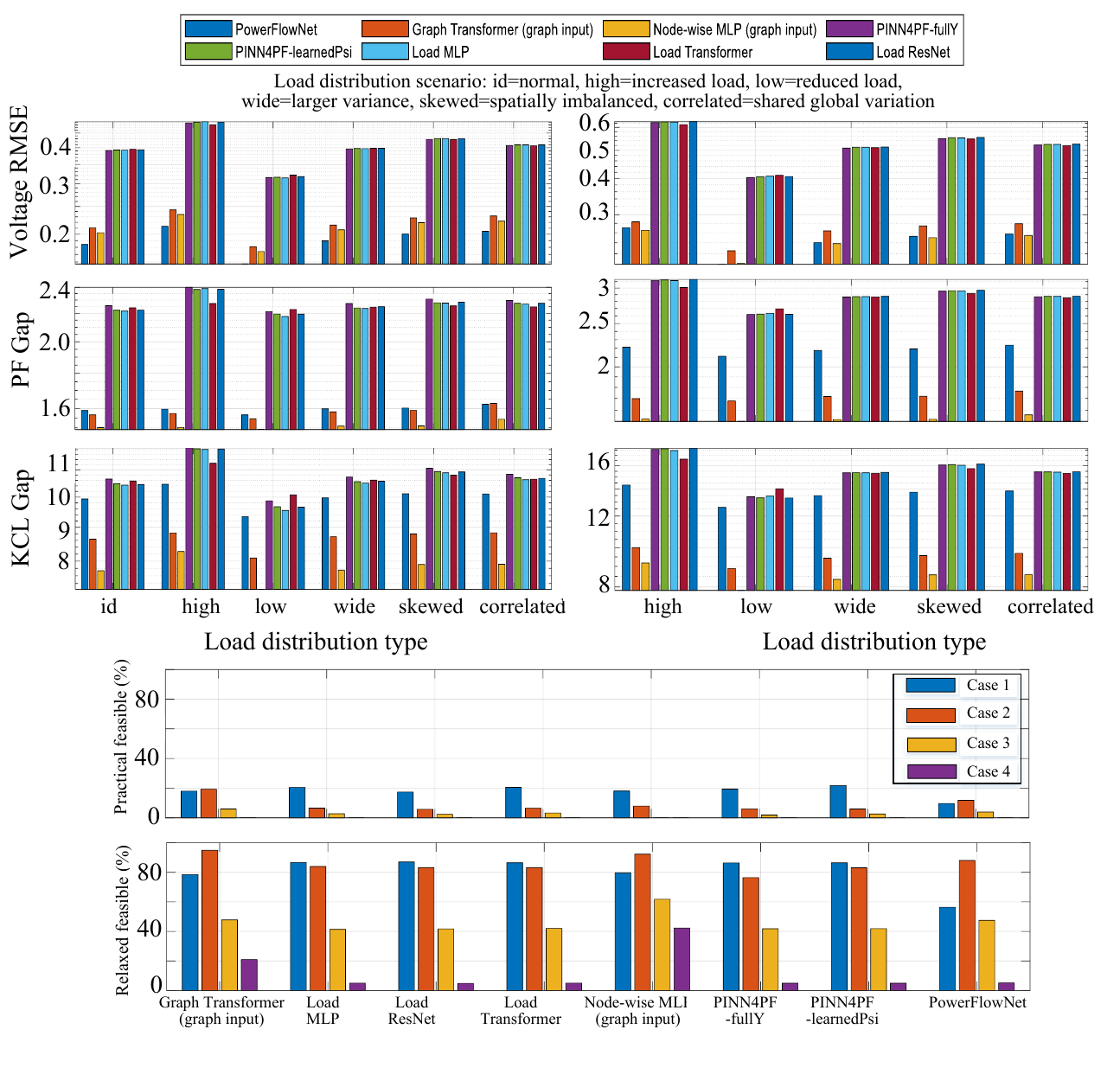}
\vspace{-0.95cm}
\caption{Gaps, practical and relaxed feasibility rates of deep learning-based PF solvers under load distribution shift and combined topology-distribution-shift conditions.}
\vspace{-0.2cm}
\label{fig3}
\vspace{-0.3cm}
\end{figure}
\subsection{Evaluation Results and Feasibility Assessment}

Simulation results in the above four cases are presented and discussed. Error and gaps curves are used to illustrate the variation of voltage RMSE, PF gap, and KCL gap with respect to topology complexity and load distribution shifts. Meanwhile, feasibility rates and scores are reported to assess whether the predicted operating points satisfy the estimated physical and operational criteria.

The voltage RMSE, $\Delta_{\mathrm{KCL}}$, and $\Delta_{\mathrm{PF}}$ are shown in Fig.~\ref{fig2}. In Cases 1, 3, and 4, a clear increasing trend can be observed when the trained PF solvers are applied to unseen topologies. The errors and gaps tend to increase as the topological deviation from the training topology becomes larger, indicating that deep learning-based PF solvers are sensitive to topology shifts even when the network size is fixed. Compared with Case 1, Case 4 exhibits stronger error and gap fluctuations across topology scenarios, suggesting that simultaneous topology and load distribution shifts further amplify the out-of-distribution effect. This result shows that the interaction between topology changes and load distribution (e.g., summer and winter datasets) changes is more challenging than topology shifts alone. In contrast, Case 2 shows relatively balanced performance across the topologies included in the training set, without a clear monotonic increase in errors or feasibility degradation with the topology ID. This is because mixed-topology training datasets enable the solvers to learn the input-output mappings associated with multiple known network configurations. However, this improvement comes at the cost of a more complex learning task and potentially longer training time. Under the same number of training epochs, mixed-topology training may still leave higher residual errors than single-topology training. Balanced performance on seen mixed topologies should therefore not be taken as evidence of high accuracy or feasibility on unseen topologies.

Feasibility rates are reported in Table~\ref{table4}, presenting the proportion of estimated operating points from the deep learning-based PF solvers that satisfy the prescribed physical and operational criteria. As shown in Table \ref{table4}, the voltage gap alone as an evaluation index is insufficient for evaluating deep learning-based PF solvers. In Cases 1 and 2, relaxed feasibility rates are relatively high for most models, whereas practical feasibility rates remain much lower, indicating that many predictions are close to the reference voltage solution but still fail to satisfy stricter line power flow consistency requirements. Mixed-topology training datasets in Case 2 improve relaxed feasibility for several solvers, with the Graph Transformer achieving the highest value of 94.9\%, but this improvement does not consistently improve practical feasibility. Under load distribution shifts in Case 3, both practical and relaxed feasibility rates decrease, showing that changes in load distributions degrade physical reliability, which is due to the intra-low generalisation of ANNs. Case 4 is the most challenging setting: all models exhibit zero practical feasibility, and only a limited fraction of samples remain relaxed feasible. Overall, Table~\ref{table4} highlights the necessity of physics-informed feasibility evaluation, since low estimation errors or high relaxed feasibility do not necessarily imply practically feasible for deploying deep learning-based PF solvers in power grids management.

\begin{table}[t]
\caption{PF FEASIBILITY RATES ACROSS THE FOUR CASE STUDIES. EACH ENTRY REPORTS PRACTICAL/RELAXED FEASIBLE RATES IN PERCENT.}    \centering
    \scriptsize
    \renewcommand{\arraystretch}{1.15}
    \begin{tabular}{p{2.5cm}|c|c|c|c}
        \hline
        \textbf{Solvers} 
        & \textbf{Case 1} 
        & \textbf{Case 2} 
        & \textbf{Case 3} 
        & \textbf{Case 4} \\
        \hline
        PowerFlowNet & 9.6/56.4 & 11.8/88.0 & 3.9/47.5 & 0.0/5.1 \\
        Graph Transformer & 18.0/78.5 & 19.3/94.9 & 5.9/47.8 & 0.0/21.0 \\
        Node-wise MLP & 18.1/79.7 & 7.9/92.3 & 0.0/61.8 & 0.0/42.2 \\
        PINN4PF-fullY & 19.4/86.4 & 6.0/76.4 & 1.9/41.8 & 0.0/4.9 \\
        PINN4PF-learnedPsi & 21.8/86.4 & 6.0/83.1 & 2.6/41.9 & 0.0/4.9 \\
        Load MLP & 20.5/86.7 & 6.5/84.0 & 2.7/41.4 & 0.0/5.0 \\
        Load Transformer & 20.6/86.4 & 6.5/83.2 & 3.2/42.0 & 0.0/5.0 \\
        Load ResNet & 17.4/87.0 & 5.7/83.2 & 2.4/41.7 & 0.0/4.9 \\
        \hline
    \end{tabular}
    \label{table4}
    \vspace{-0.4cm}
\end{table}

Fig.~\ref{fig3} presents the feasibility of the trained PF solvers under load distribution shifts and the preset feasibility criteria. Since excessively altering the load distribution range may lead to non-convergent or infeasible PF calculation, the load distribution range is intentionally kept within a moderate level to ensure that training datasets can be generated. As a result, the impact of the load distribution shifts alone is not significant. Nevertheless, the overall practical feasibility of the evaluated deep learning-based PF solvers remains low, generally below 20\%, despite their relatively good voltage-domain prediction accuracy. This indicates that accurate voltage estimation does not necessarily imply satisfaction of the underlying physical constraints or operational requirements. Among the four case studies, Case 4 exhibits the lowest feasibility across all solvers, as it combines topology shifts and load distribution shifts simultaneously. This setting is also closer to realistic power grid operation, where both network configurations and operating conditions may change over time. Therefore, the results suggest that existing deep learning-based PF solvers are still insufficiently reliable for direct deployment in practical power grid applications. Further research is required to improve their robustness, physical consistency, and operational feasibility under combined out-of-distribution conditions.

\begin{figure}
\centering
\includegraphics[width=0.5\textwidth,height=0.3\textwidth]{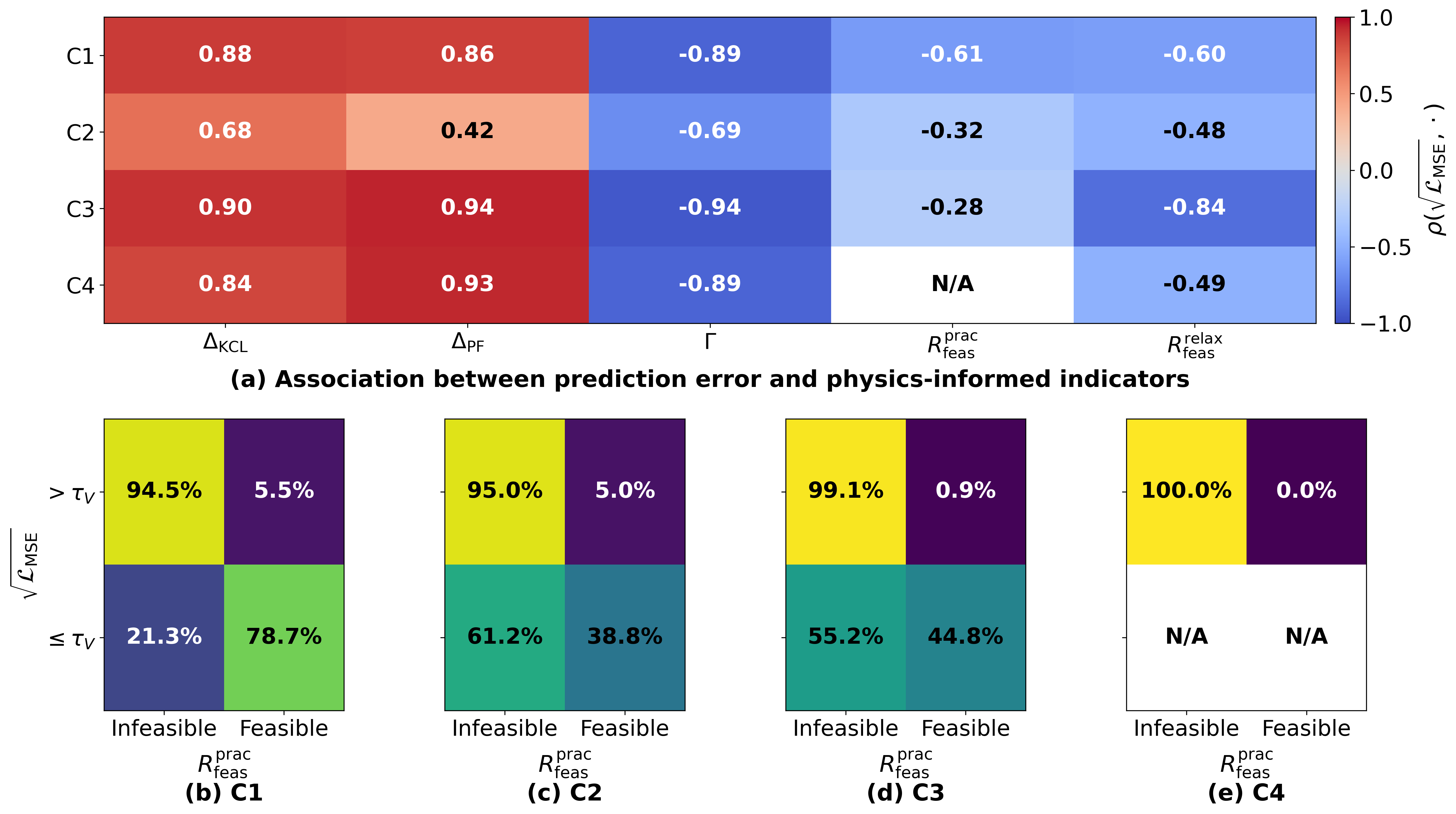}
\vspace{-0.7cm}
\caption{Relationship between voltage accuracy and physics-informed feasibility. (a) Correlations between voltage RMSE and physical indicators; (b)--(e) practical feasibility distributions for Cases 1--4.}
\vspace{-0.6cm}
\label{fig4}
\end{figure}
Fig.~\ref{fig4} further demonstrates the value of the proposed physics-informed evaluation framework. The upper panel shows that $\Delta_{\mathrm{KCL}}$, $\Delta_{\mathrm{PF}}$, and $\Gamma$ vary consistently with voltage RMSE, while providing complementary information on physical consistency. More importantly, the lower panels show that low voltage RMSE does not necessarily imply practical PF feasibility, as a considerable proportion of low-error samples remains physically infeasible. These results confirm that statistical accuracy alone is insufficient for assessing the physical admissibility of deep learning-based PF results.

Furthermore, the proposed framework can support both the development and deployment of deep learning-based PF solvers. During offline development, the proposed indicators can serve as auxiliary training objectives, model selection criteria, or calibration measures to promote physical consistency. During online operation, they can form a physics-informed acceptance mechanism that verifies KCL consistency, branch-flow deviations, voltage limits, and other operational constraints before the ANN outputs are used by downstream applications. ANN-derived operating points that fail these criteria can be rejected, recalculated using a conventional AC PF solver, or subjected to corrective procedures. This architecture provides a pathway from accurate numerical surrogates toward physically validated and trustworthy power system analysis tools while retaining the computational efficiency of deep learning-based PF solvers.

\section{Conclusion}
This paper proposed a physics-informed evaluation framework for deep learning-based power flow solvers under system shifts, including topology shifts, load-distribution shifts, and their combined effects. Beyond conventional voltage error, the framework assesses physical consistency and operational feasibility through KCL consistency, line power flow consistency, KVL residuals, and feasibility rates. Results from four case studies show that low training loss and low error do not necessarily imply high practical feasibility. Incorporating diverse topology and operating scenarios into the training set improves performance under seen conditions, but it increases training complexity and does not guarantee reliable generalization to unseen or out-of-distribution conditions. These findings indicate that deep learning-based PF solvers should be evaluated not merely as numerical regressors, but as power system analysis tools whose outputs remain physically consistent and operationally reliable. Therefore, physical feasibility and robustness to system shifts are essential for trustworthy application.

\bibliography{ref}{}
\bibliographystyle{IEEEtran}

\end{document}